# ABSORPTION AND PLASMON RESONANCE OF BI-METALLIC CORE-SHELL NANOPARTICLES ON A DIELECTRIC SUBSTRATE NEAR AN EXTERNAL TIP

Dilan Avşar[1], Hakan Ertürk[*,1], M. Pınar Mengüç[2]

[1] Boğaziçi University, Department of Mechanical Engineering,
Bebek, 34342, İstanbul Turkey
[2] Özyeğin University, Center for Energy, Environment and Economy (CEEE),
Çekmeköy, 34794, İstanbul Turkey

ABSTRACT. Absorption efficiency profiles and localized surface plasmon resonance (LSPR) wavelengths are reported for metallic core-shell nanoparticles (NPs) placed over a BK7 glass substrate. A numerical study is performed with the vectorized version of the discrete dipole approximation with surface interactions (DDA-SI-v). Gold (Au) and silver (Ag) metallic components are used for the simulations of two different core-shell structures. Absorption enhancement and the hybrid modes of plasmon resonances of the core-shell structures are compared by using a measure that defines a size configuration. It is observed that a small volume fraction of the core sizes results in shell domination over the plasmon response. An additional study is conducted to discern the sensitivity of the refractive index of nanoparticles in different surrounding environments. With a selected core-shell size configuration of Ag-Au pairs, a significant absorption enhancement with a redshift of LSPR wavelength is observed for both Ag core-Au shell and Au core-Ag shell NPs. The absorption behavior of the bare metallic NPs and selected core-shell pairs in proximity to an external probe's tip is also analyzed. The gallium phosphide (GaP) and silicon (Si) tip usage are investigated with transverse electric (TE) and transverse magnetic (TM) wave polarizations. It is observed that the dominance of light polarization on the absorption enhancement of the NPs switches at different wavelengths, where the dielectric transition for tip materials occurs. These findings show the possible targeted uses of metallic core-shell nanoparticles in several areas such as nanomanufacturing, localized heating, bio-sensing, and material detection applications.

## 1. INTRODUCTION

Metallic nanoparticles (NPs) have strong absorption behavior over the entire visible light spectrum due to their relatively large absorptive index ($k$). An increase in metal particle size generally results in a redshift and damping of dipole resonance wavelength due to reduced restoring force and decreased effects of interband transitions. Moreover, plasmon responses of the metal NPs are strongly dependent on the surrounding or neighboring medium. Hence, biological detection through optical sensing of the refractive index of the surrounding medium can effectively be performed using metal particles and their hybrid structures [1–3]. Both experimental and numerical studies have shown that metal-metal core-shell nanoparticles can have hybridized plasmon modes, which are due to bimetallic structure inducing surface and cavity plasmons depending on the core/shell ratio [3–8]. Hybrid plasmon modes of the metal-metal core-shell nanoparticles can be exploited for field enhancement, localized heating, optical detection, and controlled shell formation due to their wide tunability and the multi-peak response of localized surface plasmon resonance (LSPR) [8,9]. LSPR response of metal-metal core-shell nanoparticles may not be well interpreted from their core/shell ratio unlike the case with dielectric-metal pairs [10]. Since the plasmon peak positions depend on the LSPR wavelengths of both core and shell plasmonic materials, composition ratio and shielding of the metallic shell can affect the position and amplitude of the LSPR [3,7,10].

Zhang et al. [8] showed in their theoretical study that the bimetallic structure of core-shell NPs can have two different surface plasmon resonance (SPR) modes that lead to two peaks with ordinary and extraordinary modes. They also showed that different metal-metal arrangements can support both red-shift and blue-shift in the short (extraordinary mode) and long (ordinary mode) wavelengths as the core/shell size is varied. Since the field localizations occur both at the core-shell interface and on the outer shell surface, the size configurations of metal-metal structures can be utilized for optical detection and bio-sensing. Navas and Soni [11] experimentally and theoretically showed that metal NPs can have enhanced absorption in water compared to those in air. Moreover, they found that LSPR wavelengths red-shifts as the excitation of surface electrons needs smaller energy in medium with a larger refractive index. Comparisons of plasmon response of the metal NPs showed that Ag and Au NPs are good candidates as they have higher refractive index sensitivity, which can be utilized in bio-sensing applications.

An experimental study showed that the selective melting of metallic NPs is possible when the NPs are excited via total internal reflection (TIR) under a sharp scanning probe [12]. Localized heating and nanopatterning of an array of gold NPs on a glass substrate is further studied numerically and it is shown that exploitation of LSPR and near-field coupling of an external probe (such as an atomic force microscope (AFM) probe) with the NPs can enhance the absorption [13]. Moreover, absorption enhancement and LSPR wavelength changes are observed for the core-shell structures with dielectric and metal pairs, which can be utilized in optical tuning and localized heating applications [10,14]. It is shown that near-field coupling can enhance the absorption of gold nanoparticles placed on a dielectric substrate under a dielectric tip due to perturbation on the dielectric environment [15]. Moreover, a comparison of incident wave polarizations for the absorption enhancement of gold nanoparticles showed that TM-wave illumination creates greater field enhancement. This is attributed to the fact that TE polarization creates decaying evanescent wave transverse to the tip-NP axis, while the TM one has additional partial polarization along the tip-NP axis leading to more perturbation on the dielectric environment of the nanoparticle [14,15]. Nevertheless, the effect of a dielectric tip on the absorption of metal nanoparticles should not be generalized. Comparison of Ag and Au nanoparticles showed that Si tip damps the absorption of AgNP in contrast with the case of AuNP. The spectral absorption behavior of AgNP with Si tip results in increased damping factor dominating over the driving force. However, an increase in driving force domination results in absorption enhancement in the AuNP case [16].

The studies in the literature related to LSPR wavelength tunability and refractive index sensing are mostly considered for the NPs in free space. However, in many applications such as bio-sensing, the nanoparticles are usually placed at an interface, where they interact with light and the surrounding media. Therefore, there is a need for investigation of optical absorption behavior under varying core/shell size at an interface so that these hybrid structures can be utilized in these applications. Hence, this study focuses on Ag and Au bimetallic core-shell nanoparticles placed over a dielectric surface. Moreover, the absorption suppression of metal nanoparticles under a dielectric tip as reported in [16,17] are also applied to core-shell structures, with Si and GaP tips are analyzed considering their spectral refractive index, especially where these materials behave as a dielectric for the wavelengths greater than 490 nm [10,14,18]. Since the LSPR wavelengths of Ag and Au NPs in 50 nm diameter are around 365 nm and 515 nm [10], respectively; the damping of absorption in AgNP may not be effectively seen for longer wavelengths. Therefore, there is a need for spectral absorption analysis in a wider range of spectrum including the LSPR wavelength of both Ag and Au nanoparticles.

Nanostructures and their interaction with light and surrounding medium can be efficiently evaluated with a semi-analytical computational electromagnetic method that is referred to as discrete dipole approximation (DDA). The method was extended for configurations, where nanostructures are placed over a dielectric substrate by considering the Sommerfeld integral formulation. This approach was adopted using an open-source MATLAB™ toolbox and a new method for DDA with surface interactions (called DDA-SI) developed by Loke and Mengüç [19]. The toolbox was further improved for computational efficiency adopting a vector formulation in the vectorized version of DDA-SI (DDA-SI-v) by Talebi-Moghaddam et. al [14], where verification of the method was also provided. Recently, an updated version (called DDA-SI-z) was developed by Rostampour-Fathi et al. for the improved solver performance [20].

Considering the problem investigated in this paper, all numerical simulations are performed with DDA-SI-v. Both Ag core-Au shell (Ag@Au) and Au core-Ag shell (Au@Ag) metallic NPs are studied with their LSPR response and absorption efficiency profiles. Comparisons are done based on the plasmon responses with respect to varying size configurations and surrounding media. Refractive indices of the medium above the surface are chosen between 1 and 1.5 so that it can be applied to water, hemoglobin/plasma [21–23], and to other possible biological materials for bio-sensing applications. Further analysis will be conducted on selected core-shell pairs with the greatest absorption scenario to compare the effect of near-field coupling of external tip on the spectral absorption behavior.

## 2. PROBLEM STATEMENT AND METHOD

This study is divided into three parts for the comparison of Ag-Au core-shell metallic structures placed on a semi-infinite BK7 glass substrate subjected to an incident wave from below and within the surface with an incident angle ($\theta_i$) greater than the critical angle ($\theta_c$), resulting in a total internal reflection (TIR). Figure 1 depicts the resultant surface evanescent waves that interact with the core-shell structure defined with the core radius ($R_c$) or outer shell radius ($R_s$). As shown in the Figure 1, nanoparticle interacts with both surface evanescent wave, and the surrounding media that consist of upper medium and semi-infinite glass substrate. In this configuration, the TE-polarized wave is preferred for the investigation as the resulting absorption efficiency ($Q_{abs}$) is slightly larger than that with the TM-polarized wave. The incident TE-polarized wave is in the range of 300 nm to 800 nm. It should be noted that the incident wavelength in this problem is defined for the vacuum. However, the wavelength will be contracted as it travels into the dielectric substrate. Then, the interaction with the nanoparticle will take place in the upper-medium, where the wavelength is calculated again for the vacuum or other dielectric upper-medium cases. Although the wavelength undergoes a change in these media, the frequency remains the same due to the nature of the light. The first numerical analysis is based on the effect of size configuration on the absorption efficiency ($Q_{abs}$) and LSPR wavelength ($\lambda_{max}$) of the core-shell structure being placed in a vacuum on top of the substrate.

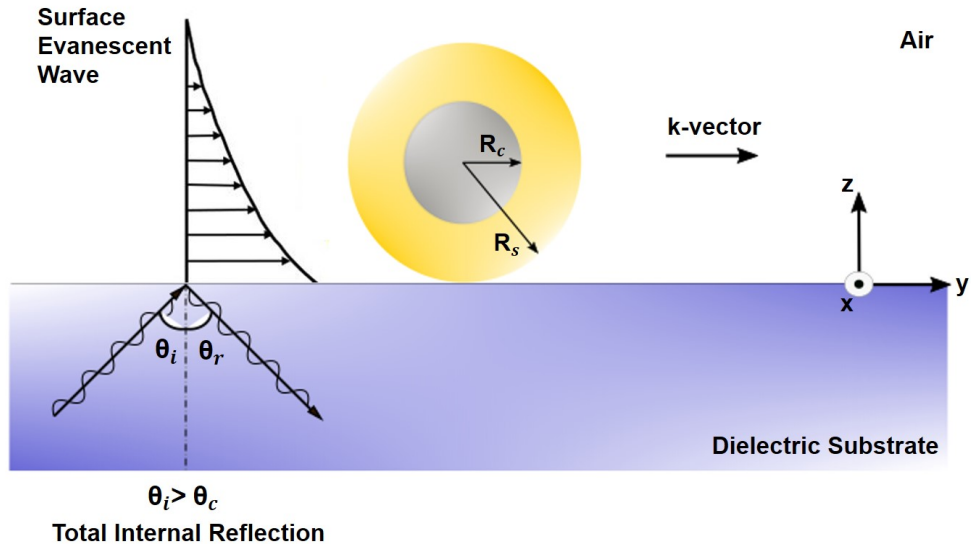

Figure 1. A schematic view of the configuration for a single core-shell system being represented with core radius ($R_c$) and outer shell radius ($R_s$). TE-polarized wave is incident with an angle greater than the critical angle ($\theta_c$) for TIR under the BK7 substrate that creates surface evanescent wave propagating and decaying in y- and z-directions, respectively

The size effect is studied for fixed core (FC) and fixed nanoparticle size (FNP) cases for Ag core-Au shell (Ag@Au) and Au core-Ag shell (Au@Ag) core-shell NPs. In the FC case, outer shell diameter is increased gradually to 50 nm while the metallic core is kept at 20 nm in diameter. The FNP cases involve core diameters between 15 nm and 45 nm as the overall NP size is fixed at 50 nm in diameter. The size configurations for the FNP and FC cases are defined for core and outer shell diameters with 3 nm and 2 nm increments, respectively, and are presented in Table 1. Moreover, the change in core-shell size is parameterized with the volumetric filling ratio ($f_r$) of the core material, which is defined as the ratio of the core volume based on core diameter, $D_c$, to the nanoparticle volume defined based on outer shell diameter, $D_s$, as $f_r=(D_c/D_s)^3$. The $f_r$ range considered is presented in Table 1.

The second part of the problem considers the refractive index sensitivity of these structures as the dielectric medium over the BK7 substrate is varied. The effect of the top surrounding medium on the $Q_{abs}$ and $\lambda_{max}$ of the metallic core-shell NPs is studied with only FC cases. The core-shell pairs are considered with shell thicknesses, $t_s$, between 5 nm and 15 nm, while the core is kept at 20 nm diameter. Since the refractive index of the BK7 substrate is about 1.52, upper medium refractive indices are varied between 1 and 1.5 so that the incident angle between $\theta_c$ and 90° can support TIR to create surface evanescent wave above the substrate. The selection of incident angle is important for obtaining the optimum absorption efficiency profiles. For the analysis, the intensity profile of the surface evanescent wave is considered as shown in the Eqs. (1) and (2), which depend on wave vector, $k$, incident angle, $\theta_i$, critical angle, $\theta_c$, elevation from surface, $z$, speed of light, $c$, wavelength, $\lambda$, and refractive indices of top medium and the substrate, $n_m$ and $n_{BK7}$, respectively.

$$I = I_0 \exp\left\{-k\left[\left(\frac{\sin\theta_i}{\sin\theta_c}\right)^2 - 1\right]^{1/2} z\right\} \quad (1)$$

$$k = \frac{c}{\lambda}\frac{n_m}{n_{BK7}} \quad (2)$$

As shown in equation (1), it should be noted that increasing angle of incidence will result in a decrease in surface evanescent wave intensity, which also decreases the coupling between the wave and the nanoparticle, and hence, the absorption. Therefore, the incident angles are selected as close as possible to the critical angles for each dielectric medium case. Table 2 shows the considered medium refractive indices ($n_m$) with their corresponding incident angles ($\theta_i$), which are greater than the critical angles ($\theta_c$) defined by Snell's law of refraction.

The last part of the problem is designated to consider the maximum absorption efficiency cases of Ag@Au and Au@Ag core-shell nanoparticles with fixed nanoparticle size found in the first part. The effect of an external tip on the LSPR response of the nanoparticles is investigated with the selected pairs. In addition to the TE-polarized wave in the first and second parts, the TM-polarized case is also considered in this part due to its partial polarization along the tip-NP axis as stated in [14–16]. The study in [16] showed that after 600 nm, the absorption level of the AuNP is converged to minimum values with negligible changes. However, the wavelength range of 300 nm-450 nm in [16] and [17] was not enough for the AgNP to experience the effects of the dielectric transition of the tip material. Therefore, the incident wavelength in this part is considered in 300 nm-600 nm range to examine the LSPR response of both bare metal and core-shell NPs. As shown in Figure 2, the position of the spherical NP is considered as the origin of the cartesian coordinate system, and the tip is positioned over the NP at a coordinate (0, 0, 52) nm. The geometry of the external tip is defined with 10 nm tip radius, $r_{tip}$, 50 nm shaft diameter, $D_{shaft}$, so that the cone angle, $\alpha$, is 16.7°. Both TE and TM polarized incident waves are considered; where TE-polarized wave creates the evanescent wave transverse to the tip-NP axis in y-direction, whereas TM-polarized wave has both y and z components.

Table 1. Core and shell sizes for FC and FNP cases with corresponding volumetric filling ratio ($f_r$) ranges.

| NP Type | Core Diameter, $D_c$ (nm) | Shell Diameter, $D_s$ (nm) | $f_r$ Range |
|---|---|---|---|
| Fixed Core (FC) | 20 | 22 – 50 | 0.751 – 0.064 |
| Fixed NP (FNP) | 15 – 45 | 50 | 0.027 – 0.729 |

Table 2. Refractive indices of the medium ($n_m$) with their critical ($\theta_c$) and incident angles ($\theta_i$).

| $n_m$ | 1 | 1.1 | 1.2 | 1.25 | 1.3 | 1.4 | 1.5 |
|---|---|---|---|---|---|---|---|
| $\theta_c$ | 41.13° | 46.34° | 52.11° | 55.30° | 58.76° | 67.04° | 80.58° |
| $\theta_i$ | 45° | 50° | 55° | 60° | 60° | 70° | 85° |

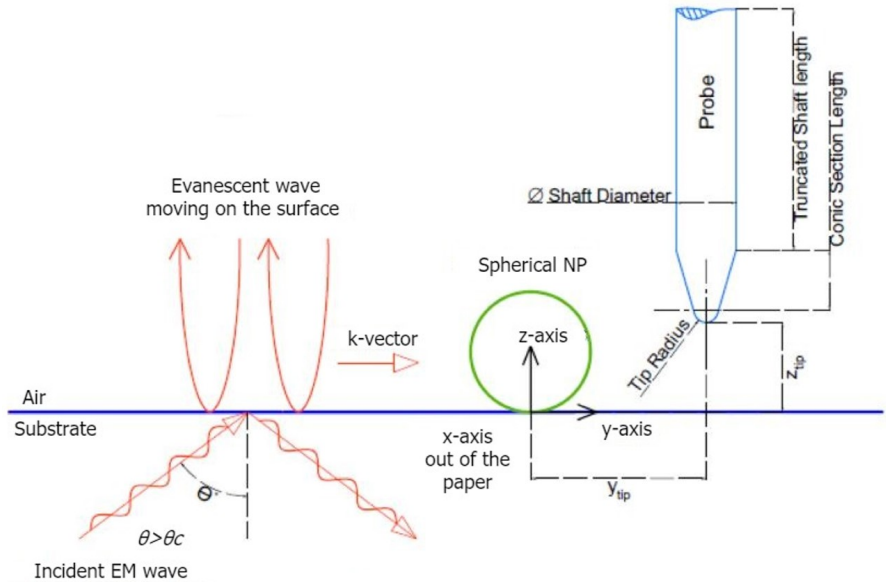

Figure 2. A cross-sectional view of the external tip and nanoparticle configuration [14].

A MATLAB toolbox, the vectorized version of DDA-SI (DDA-SI-v), is utilized for the spectral absorption efficiency calculations in this study [14]. For the first part of the problem with single core-shell nanoparticles, the accurate calculation of absorption efficiency necessitates a sufficient number of dipoles ($N_{NP}$) to represent the nanoparticles. Hence, an optimum dipole number is determined as $N_{NP}$=1472 based on the converged value of maximum $Q_{abs}$ values, which has been considered in the previous study of dielectric-metal core-shell pairs [10]. However, for the external tip and NP interactions, $N_{NP}$ is decreased to 552 due to a significant increase in the required computation time and memory with the introduction of the external probe. Then, a number of dipoles for the external probe, $N_{Probe}$, is analyzed for the optimum shaft length so that a further increase in the probe length does not change the absorption significantly. In this study, the optimal shaft length is considered when the relative change of absorption becomes within 5%. Similar design has been used for the single AuNP and $SiO_2$ core-Au shell cases with the truncated shaft length of 390 nm in [13,14].

Throughout this study, the spectral refractive indices of the Ag and Au metals are obtained from the experimental measurements reported in [24], while an empirical relation of refractive index with wavelength based on Sellmeier equation is used for the BK7 substrate. For the tip materials, the data for GaP and Silicon (Si) are used from [25,26], respectively.

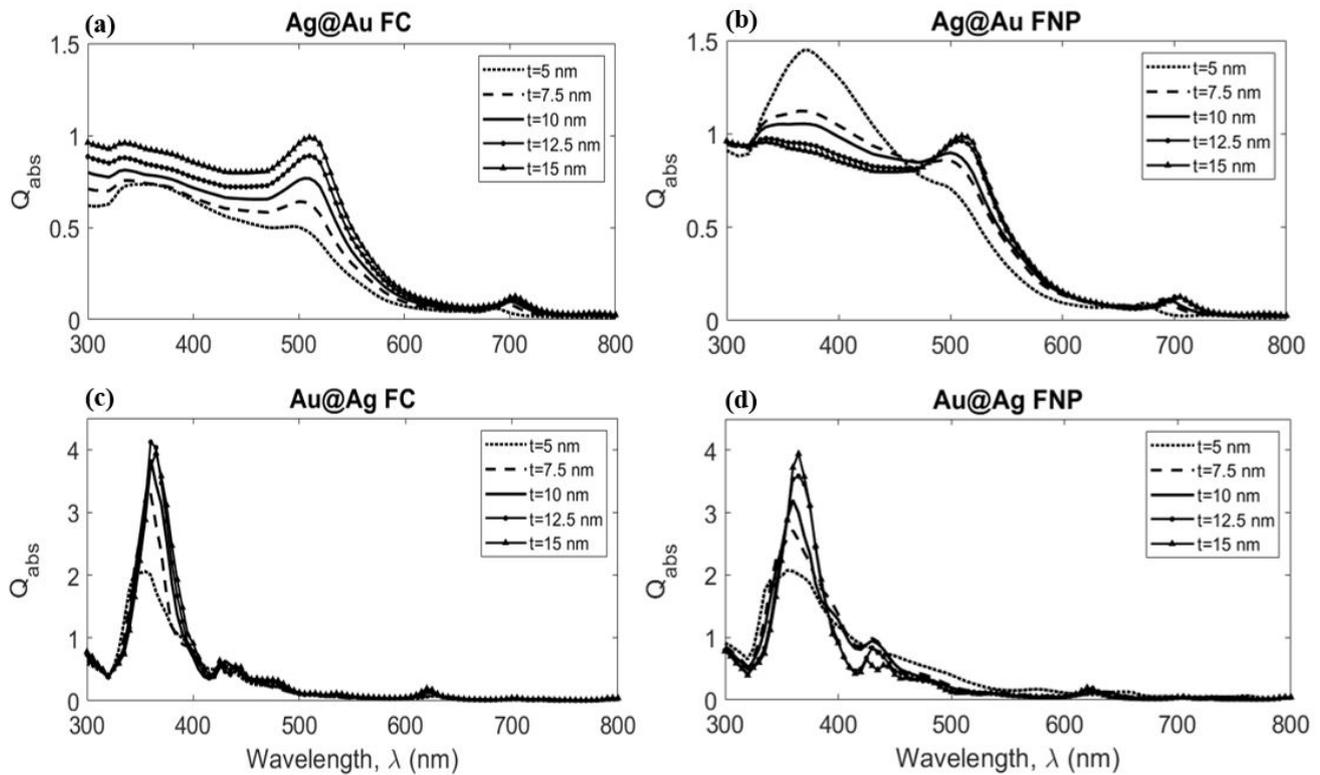

Figure 3. Spectral absorption efficiency, $Q_{abs}$, of Ag core-Au shell (a-b) and Au core-Ag shell (c-d), represented for FC (a,c) and FNP (b,d) cases.

## 3. RESULTS AND DISCUSSIONS

### 3.1 Fixed core and fixed nanoparticle size

The plasmon responses of metal-metal core-shell NPs are numerically analyzed considering their spectral absorption efficiency values for both FC and FNP cases as shown in Figure 3. Plasmon mode hybridizations are observed in FC and FNP cases of Ag core-Au shell (Ag@Au) NPs with varying shell thickness, $t$, between 5 nm and 15 nm. An increase in Au shell thickness results in a redshift of the ordinary modes (secondary peaks) in longer wavelengths that are located around 515 nm as shown in Figures 3 (a) and (b). These modes correspond to the LSPR wavelength of bare Au NP, which explains the Au shell domination as its thickness increases. Even if the extraordinary modes (first peaks) in Figure 3 (a) is not clear for shell thicknesses greater than 5 nm, an increase in shell thickness blueshifts this mode. This mode of plasmon resonance corresponds to field enhancements at the core-shell interface, while the ordinary modes are the result of field localization at the outer surface of the Au shell. The extraordinary and the ordinary modes, which are shown in Figures 3 (a) and (b), are attributed to the surface plasmon resonances at the Ag core-Au shell interface, and at the Au shell-medium interface, respectively.

Contrary to the Ag@Au case, the spectral absorption profiles of Au core-Ag shell (Au@Ag) case with varying shell thicknesses have the greatest absorption peak at the extraordinary modes. Figures 3 (c) and (d) show the shielding effect of Ag shell on the plasmon response of the core-shell NP with narrower LSPR wavelength bandwidth. LSPR wavelengths redshifts by 10 nm to 365 nm, which corresponds to bare Ag NP's plasmon wavelength, as the Ag shell thickness increases for both FC and FNP cases. Ordinary modes (secondary peaks) are observed around 420 nm, with decreasing $Q_{abs}$ values in Figure 3 (d), for the Ag shell thickness greater than 5 nm. In this scenario, the Ag shell dominates over the Au core with field localizations at the outer shell surface for both extraordinary and ordinary modes of the LSPR wavelength. Therefore, the prominent peaks are only observed in the extraordinary modes that arises from the resonance at the Ag shell-medium interface.

In Figure 4, the effects of size configuration on the LSPR response of the NPs are further analyzed with the contour plots of FC cases and the volumetric filling ratio of the core materials based on the size configurations summarized in Table 1. As discussed above, Ag@Au core-shell NP can support plasmon hybridization, which is depicted in Figure 4 (a) with two peak regions located around 365 nm and 515 nm. For outer shell diameters smaller than 25 nm, further decrease in shell thickness ($t < 5$ nm) results in greater $Q_{abs}$ values, which cannot be observed in Figure 3 (a). Further Au shell thickness increase in this FC case dominates the ordinary mode so that its LSPR wavelength moves towards that of the bare Au NP (515 nm). When the volumetric filling ratio, $f_r$, of the core is smaller than 0.2, the absorption enhancement is not observed as shown in Figure 4 (b). However, a further increase in $f_r$ beyond 0.2 leads to increasing maximum $Q_{abs}$ values of Ag@Au NP by 22% and 91% for FC and FNP cases, respectively. When the $f_r$ reaches its highest value, the FNP case has greater absorption than the one in the FC case since the FNP represents greater Ag core size (45 nm) than the FC case with 20 nm core diameter. As the $f_r$ of the Ag core increases, the LSPR wavelength of the Ag@Au NP stabilizes around the bare Ag NP's LSRP wavelength. Similar to spectral absorption efficiency patterns shown in Figures 3 (c) and (d), the Ag shell dominates the LSPR wavelengths around 365 nm for the cases with $f_r$ of the Au core smaller than 0.2, as shown in Figures 4 (c) and (d). Results for Au@Ag NP pair shows that further increase in $f_r$ results in blueshift of the LSPR wavelength towards 335 nm as its maximum $Q_{abs}$ values decrease by 280% and 135% for the FC and FNP cases, respectively. By comparing the amount of damping in the $Q_{abs}$ values and shifts in the LSPR wavelengths, metal-metal core-shell NPs can be optically detected for material prediction and core/shell formation.

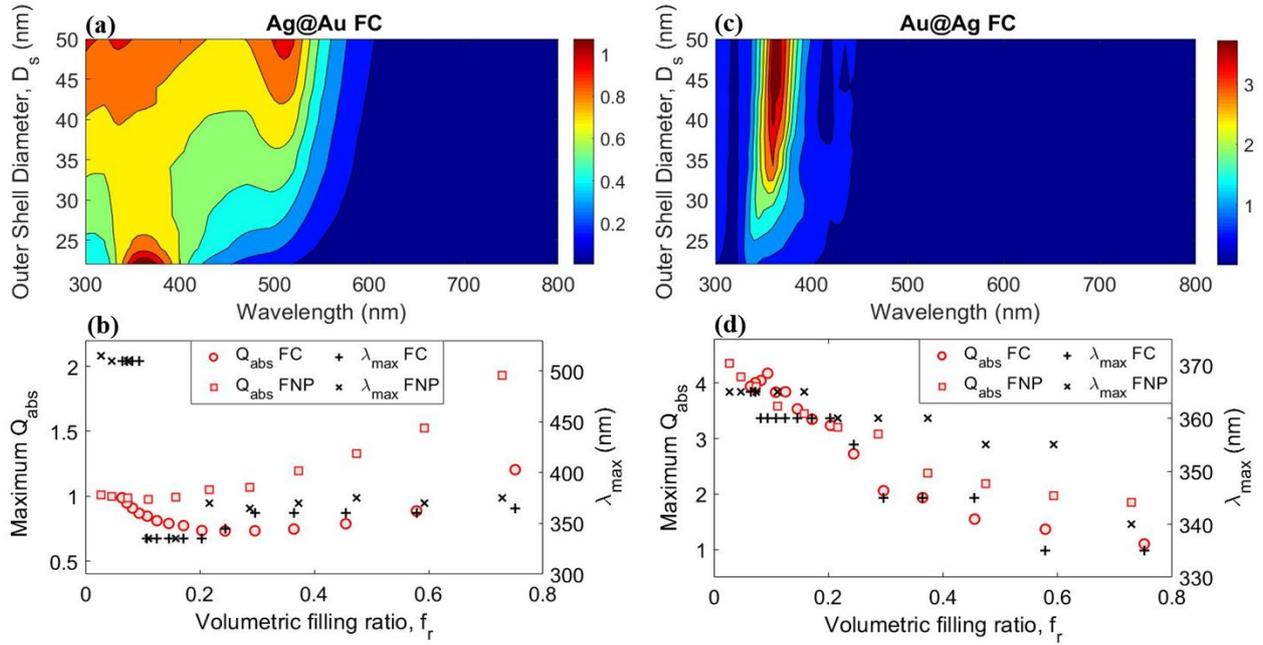

Figure 4. Ag core-Au shell (a-b) and Au core-Ag shell (c-d): Contour plots of spectral absorption efficiency, $Q_{abs}$, for FC (a,c). Maximum $Q_{abs}$ and LSPR wavelengths, $\lambda_{max}$, with respect to volumetric filling ratio, $f_r$, for both FC and FNP cases (b,d).

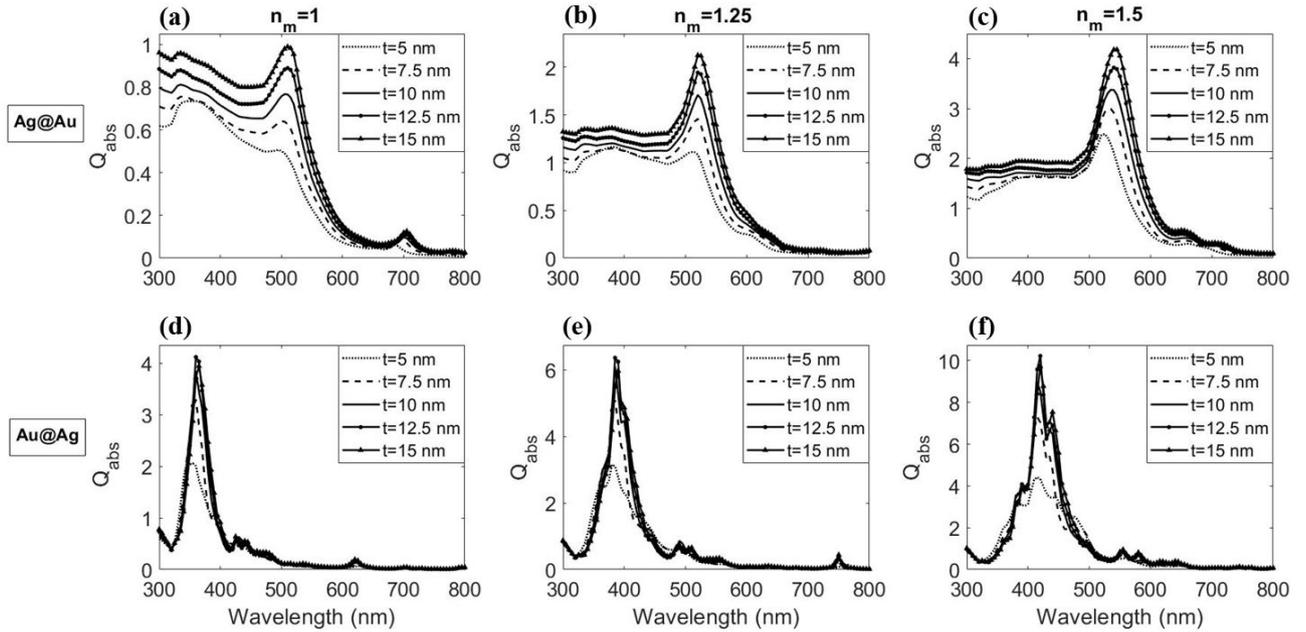

Figure 5. Spectral absorption efficiency, $Q_{abs}$, for Ag@Au (a-c) and Au@Ag (d-f) in mediums with $n_m$= 1 (a,d), 1.25 (b,e), and 1.5 (c,f).

## 3.2 Refractive index sensitivity

The effect of upper-medium on the plasmon response of the metal-metal core-shell NPs on the substrate is analyzed for the FC cases with 5 different shell thicknesses in order to include the shell formation effect. Figure 5 depicts the spectral $Q_{abs}$ values with three different refractive indices of the top medium, $n_m$, for both Ag@Au and Au@Ag NPs. The corresponding incident angles of the TE-polarized wave are chosen from Table 2 for each medium refractive index. Hybridization of plasmon modes diminishes as the refractive index of upper-medium increases for the Ag@Au core-shell case as shown in Figures 5 (a-c). Compared to the vacuum case ($n_m$=1) in Figure 5 (a), increasing $n_m$ leads to greater $Q_{abs}$ values with redshifts in LSPR wavelengths from 510 nm to 540 nm, for maximum shell thickness case. Here, in Figures 5 (a-c), an increase in $n_m$ also leads to broadening and narrowing of extraordinary and ordinary LSPR modes, respectively. In this scenario, the interaction between the Au shell and the dielectric medium dominates the one at core-shell interface. Hence, the only plasmon response of Ag@Au NP is observed for the ordinary mode. Similar LSPR wavelength redshifts and absorption enhancement can be observed in the Au@Ag case with an increase in $n_m$. However, the spectral $Q_{abs}$ values of the Au@Ag case in Figures 5 (d-f) show the opposite trend of plasmon peak formation of the Ag@Au case. While the vacuum case provides one LSPR wavelength around 365 nm, an increase in $n_m$ causes distinguishable ordinary modes at longer wavelengths around 415 nm. These secondary peaks of the LSPR wavelengths start to move away from the first peaks as the refractive index of the medium is increased. Moreover, both Ag@Au and Au@Ag cases have enhanced absorption due to an increase in both nanoparticle size and refractive index of the medium above the substrate.

Further analysis is carried out with $n_m$ values between 1 and 1.5 for the selected Ag@Au and Au@Ag NPs with $D_c$=20 nm and $D_s$=50 nm. As discussed above, increasing $n_m$ values give rise to redshifts in LSPR wavelengths with an increase of $Q_{abs}$ values. For the selected size configurations, numerical simulations are performed with finer wavelength ranges, which are around the previously evaluated LSPR wavelengths. Figure 6 shows the maximum $Q_{abs}$ and corresponding LSPR wavelengths as a function of $n_m$ for both metal-metal pairs. A gradual increase in the refractive index of the medium from 1 to 1.5 shows that LSPR wavelength of the Ag@Au NP redshifts by 32 nm with absorption enhancement by more than 320%. In the Au@Ag case, this enhancement is observed as 140% with redshift in LSPR wavelength by 57 nm. This comparison shows that there is a balance between the tunability range of the LSPR wavelength and the absorption enhancement, when shell materials are switched.

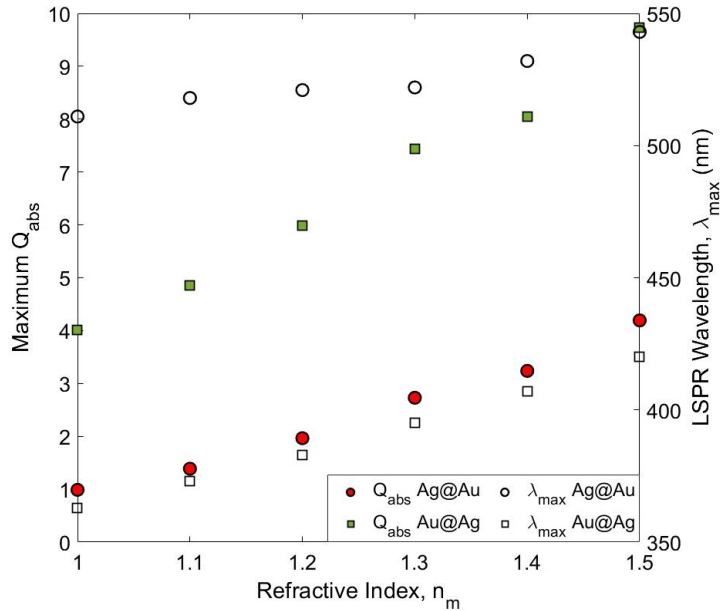

Figure 6. Maximum $Q_{abs}$ and LSPR wavelengths, $\lambda_{max}$, with respect to medium refractive indices, $n_m$ for both Ag@Au and Au@Ag.

Considering the Ag@Au case with maximum absorption enhancement shown in Figure 6, normalized field intensity plots are studied to understand the shell-medium interaction. Figures 7 (a-f) show the field localizations observed in the Ag@Au pair for the medium refractive indices, $n_m$, between 1 and 1.5. It is found that field localizations at the Ag core starts to diminish as the $n_m$ values increase from 1 to 1.5. Meanwhile, an increase in shell-medium interaction leads an enlargement of field localizations on the outer surface of the Au shell with multiple peak points. These multiple peak points reflect the surface charge density oscillations changing with the dielectric medium at corresponding wavelengths. As the refractive index of the dielectric medium increases, the local maxima points on the surface are observed to be evenly distributed showing that uniform heating in the medium is possible. This case also explains the effect of shell-medium interaction on the redshifted LSPR wavelengths that are found close to the LSPR wavelength of bare AuNP.

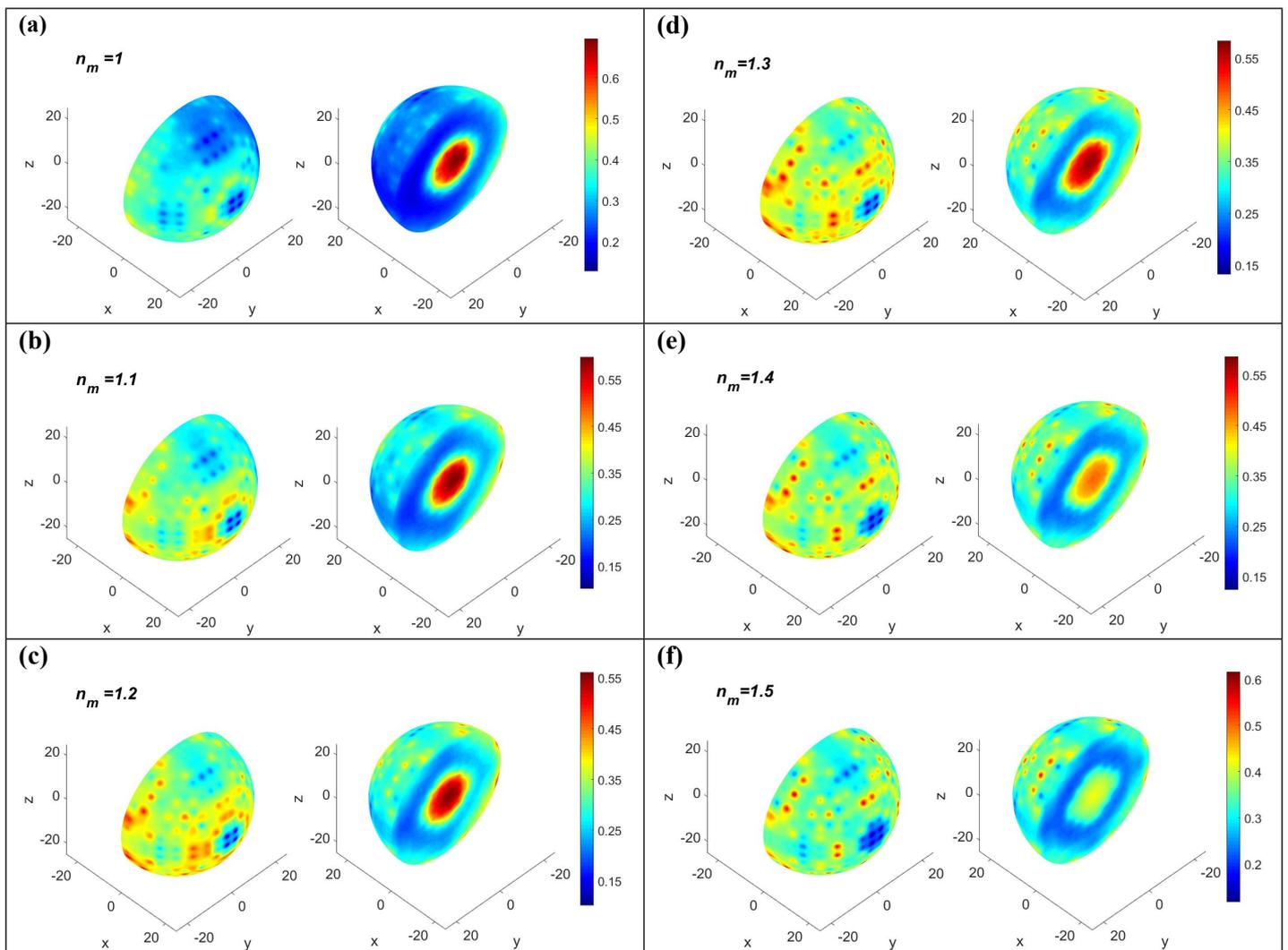

Figure 7. Normalized field intensity plots of Ag@Au medium refractive indices, $n_m$, varied between 1-1.5: (a) $n_m = 1$, $Q_{abs}= 0.99$ at $\lambda_{max}= 511$ nm, (b) $n_m = 1.1$, $Q_{abs}= 1.39$ at $\lambda_{max}= 518$ nm, (c) $n_m = 1.2$, $Q_{abs}= 1.97$ at $\lambda_{max}= 521$ nm, (d) $n_m = 1.3$, $Q_{abs}= 2.73$ at $\lambda_{max}= 522$ nm, (e) $n_m = 1.4$, $Q_{abs}= 3.24$ at $\lambda_{max}= 532$ nm, (f) $n_m = 1.5$, $Q_{abs}= 4.19$ at $\lambda_{max}= 543$ nm.

### 3.3 External tip and nanoparticle interaction

As discussed in Section 2, an optimum shaft length is analyzed with the converged values of the absorption efficiency of the metal nanoparticle. One case is evaluated for AuNP in 50 nm diameter illuminated by TM-polarized wave at $\lambda$=520 nm. The external probe is located above the NP at (0, 0, 52) nm and made of silicon with the shaft length varying between 150 nm and 810 nm while the nanoparticle is represented with $N_{NP}$=552 rather than 1472. As shown in Figure 8, changes in external probe's shaft length results in oscillating absorption efficiency values that converge to $Q_{abs}$=1.74. Considering a 5% relative error band, optimum truncated shaft length is found as $L_{shaft}$=470 nm with the corresponding number of dipoles for the external probe, $N_{Probe}$=8083. It should be noted that the convergence analysis is also carried out for other cases such as TE-polarization, GaP tip, and different wavelengths, which are found still within the 5% relative error band. However, the analyses for the Ag NP case or other core-shell NPs are not shown in this study.

The effect of external tip material is compared with the spectral absorption efficiency values of core-shell and bare metallic nanoparticles for the wavelengths between 300 nm and 600 nm. The core-shell pairs are selected from the optimum size configurations found in Section 3.1 with Ag@Au ($D_c$=45 nm, $D_s$=50 nm) and Au@Ag ($D_c$=15 nm, $D_s$=50 nm). Predicted $Q_{abs}$ values of the bare AgNP and AuNP with $D$=50 nm are taken as reference for the comparisons. An example of tip material effect is shown in Figure 9 for bare metallic nanoparticles illuminated with TE and TM polarized waves. The results show that TE-polarized cases represent negligible variation in maximum $Q_{abs}$ values for both AgNP and AuNP under GaP and Si tip. Since TE-polarized wave is transverse to the tip-NP axis, the spectral $Q_{abs}$ values follow the ones found for bare metallic NP's only with damping. The damping effect of external tip introduction is summarized for all NP types including the core-shell NPs in Table 3, where the maximum absorption efficiency values found in no tip cases are decreased when GaP and Si tips are used in TE-polarized case.

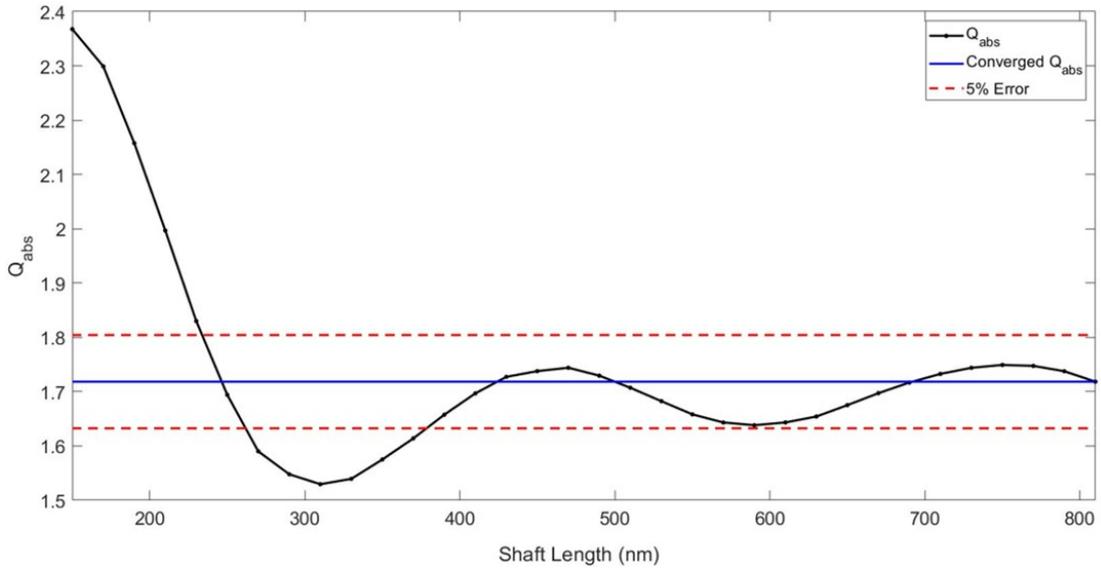

Figure 8. Converged absorption efficiency values of AuNP with changes in external probe shaft length.

**Table 3.** Summary of the maximum absorption efficiency values of the nanoparticles with their LSPR peak wavelengths represented as $Q_{abs}$ @ $\lambda_{max}$. (Table 3 – revised manuscript)

| NP type ($D$=50 nm) | Polarization | $Q_{abs}$ @ $\lambda_{max}$ | | |
|---|---|---|---|---|
| | | w/o tip | GaP tip | Si tip |
| AuNP | TE | 1.07 @ 515 nm | 1.02 @ 516 nm | 1.02 @ 516 nm |
| | TM | 1.05 @ 515 nm | 1.75 @ 521 nm | 1.75 @ 521 nm |
| AgNP | TE | 4.95 @ 370 nm | 5.01 @ 372 nm | 4.99 @ 372 nm |
| | TM | 4.62 @ 370 nm | 3.51 @ 373 nm | 3.06 @ 373 nm |
| | | 0.95 @ 490 nm | 1.79 @ 491 nm | 1.59 @ 491 nm |
| Ag@Au ($D_c$=45 nm) | TE | 1.94 @ 377 nm | 1.70 @ 380 nm | 1.69 @ 380 nm |
| | TM | 1.86 @ 376 nm | 2.03 @ 385 nm | 1.96 @ 390 nm |
| Au@Ag ($D_c$=15 nm) | TE | 4.39 @ 364 nm | 5.00 @ 372 nm | 4.81 @ 370 nm |
| | TM | 4.01 @ 364 nm | 3.53 @ 373 nm | 3.01 @ 370 nm |
| | | 0.93 @ 490 nm | 1.61 @ 490 nm | 1.52 @ 490 nm |

Considering TM-polarized cases of AgNP under external tip in Figure 9 (a), the spectral absorption behavior shows that the use of GaP tip results in maximum $Q_{abs}$ values 13% at the 1$^{st}$ peak and 11% at the 2$^{nd}$ peak greater than the ones found with the use of Si tip. The corresponding LSPR wavelength, $\lambda_{max}$, for these two peaks are 373 nm and 491 nm, respectively. Since both tip material behaves as a perfectly dielectric after $\lambda$=490 nm, absorbing behavior of the tips creates near-field interactions damping the absorption of AgNP at $\lambda$=373 nm. This effect can also be seen in Figure 9 (b) for AuNP with TM-polarization, where there is an oscillatory behavior of $Q_{abs}$ values for the wavelengths smaller than 490 nm. Therefore, the variation in maximum $Q_{abs}$ values is negligible, below 1%, for both tip materials as they behave perfectly dielectric at the peak wavelength ($\lambda_{max}$=521 nm) of the AuNP case.

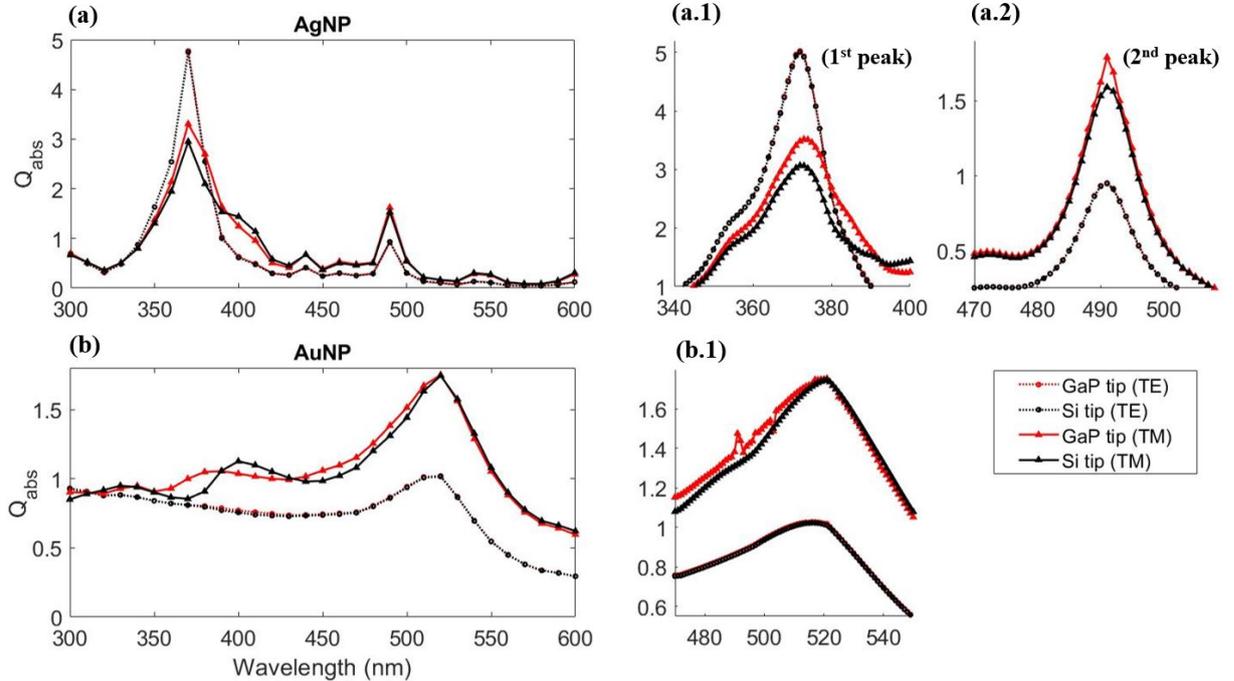

Figure 9. Spectral absorption efficiency, $Q_{abs}$, and close-up plots of peak regions of AgNP (a, a.1, a.2) and AuNP (b, b.1) under the effect of GaP and Si tips with TE/TM-polarizations.

Similar to bare metal cases, GaP tip is found to be more effective than Si tip for absorption enhancement of the core-shell NPs, as listed in Table 3. Hence, further comparisons for the spectral absorption behavior of the NPs will be based on the GaP tip. The comparison of the selected core-shell pairs with TE and TM polarized cases for the absorption enhancement is considered next. It is shown in Figure 10 (a) that the spectral absorption pattern of Au@Ag does not change significantly for TE and TM-polarized cases when there is no tip. Absorption pattern similarity is also observed for TE case with and without GaP tip for which the LSPR wavelengths differ by less than 1%. Moreover, as the TE-polarization direction is transverse to the tip-NP axis, external tip does not alter the field localizations of the NP as shown in Figures 11 (a,b). It is found that the use of GaP tip increases the maximum absorption efficiency by 14% in the TE-polarized case. For the TM-polarization cases, the maximum $Q_{abs}$ of the Au@Ag NP is damped by 12% around $\lambda$=370 nm. This kind of damping and enhancement of the absorption efficiency also exist in the bare AgNP case shown in Figure 9 (a). The main reason behind this scheme is the Ag shell domination for the selected Au@Ag pair, which can be seen in Figure 11, where the local field intensities are collected mostly on the outer shell of the NP. Moreover, the dielectric transition of the GaP tip can be observed in Figures 11 (e) and (f) with the TM-polarized case. It is shown that the field intensity is localized on top of the NP for the 1st peak wavelength, where the imaginary part of the refractive index, $k$, of GaP is nonzero in Figure 11 (e). However, as shown in Figure 11 (f), field localizations move away from the apex of the NP and they are distributed over the Ag shell at the 2nd peak wavelength, where the damping effect of GaP tip no longer exists.

It is observed in Figures 11 (a-b) that TE-polarized cases with and without external tip for a given wavelength have similar field localizations over Au@Ag NP as there is no y component of the polarization. Similarly, for the Ag@Au NP with and without a tip in Figures 12 (a,b), the field is localized both at the Ag core and on the outer surface of the Au shell. Considering Figure 10 (b), the major effect on the absorption is observed for the GaP tip with the TM-polarized wave case. Contrary to the Au@Ag case, the TM-polarized case with GaP tip results in 10% absorption enhancement of Ag@Au around $\lambda$=390 nm. This can be attributed to the absorption enhancement of AuNP with external tips in TM-polarization cases in Table 3. When the driving force dominates over the damping due to Au shell material, absorption enhancement of Ag@Au NP is observed more than 65%. Therefore, Ag@Au NP has also similar behavior of bare AuNP as its Au shell dominates over Ag core material. Moreover, field intensity plots of Ag@Au in Figures 12 (c), (d) show the effect of external tip placement over the NP as the field is collected from the -y side of the outer shell surface to the apex of the NP, which proves the vertical coupling between the tip and the NP. As discussed in the subsection 3.2, it should also be noted that spotted field localizations on the outer surface of the metallic shell result from the surface charges at different modes of oscillations. Figures 11 and 12 represent these local maxima points regarding the unique conditions of given cases. Moreover, the nature of the dipole representation of the nanoparticles throughout the numerical analysis leads to a dotted texture on the field intensity plots.

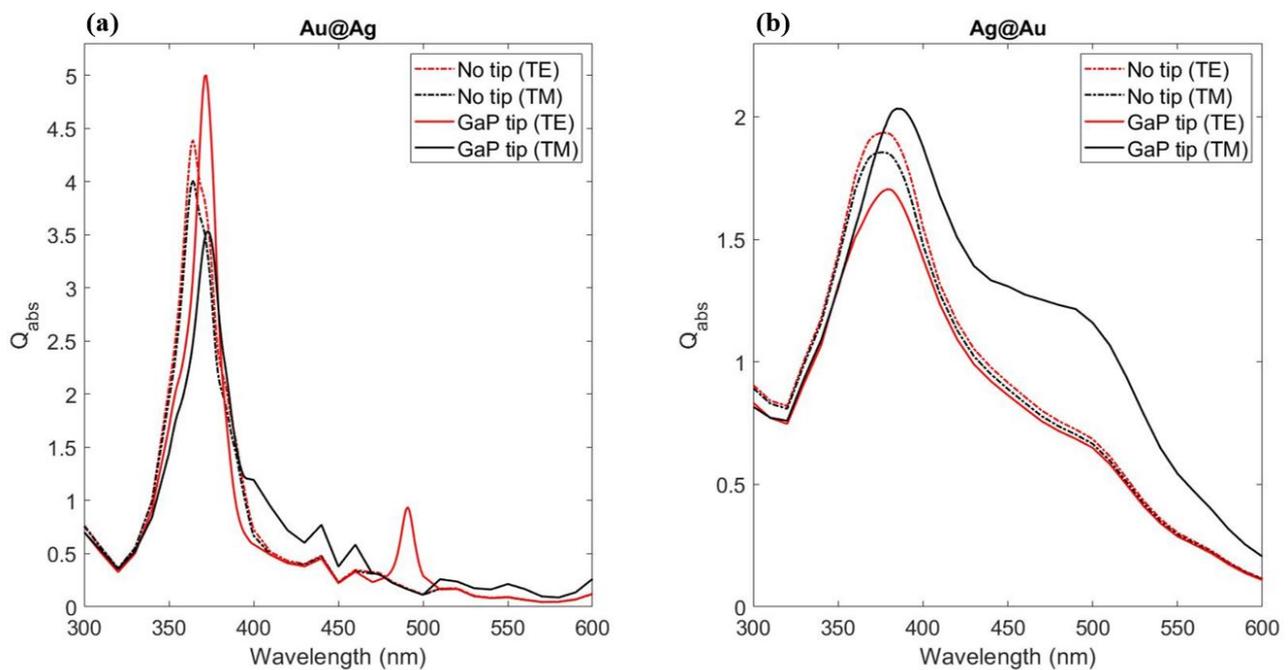

Figure 10. Spectral absorption efficiency, $Q_{abs}$, of core-shell pairs: Ag@Au (a) and Au@Ag (b) with and without using GaP tip in TE/TM-polarization cases.

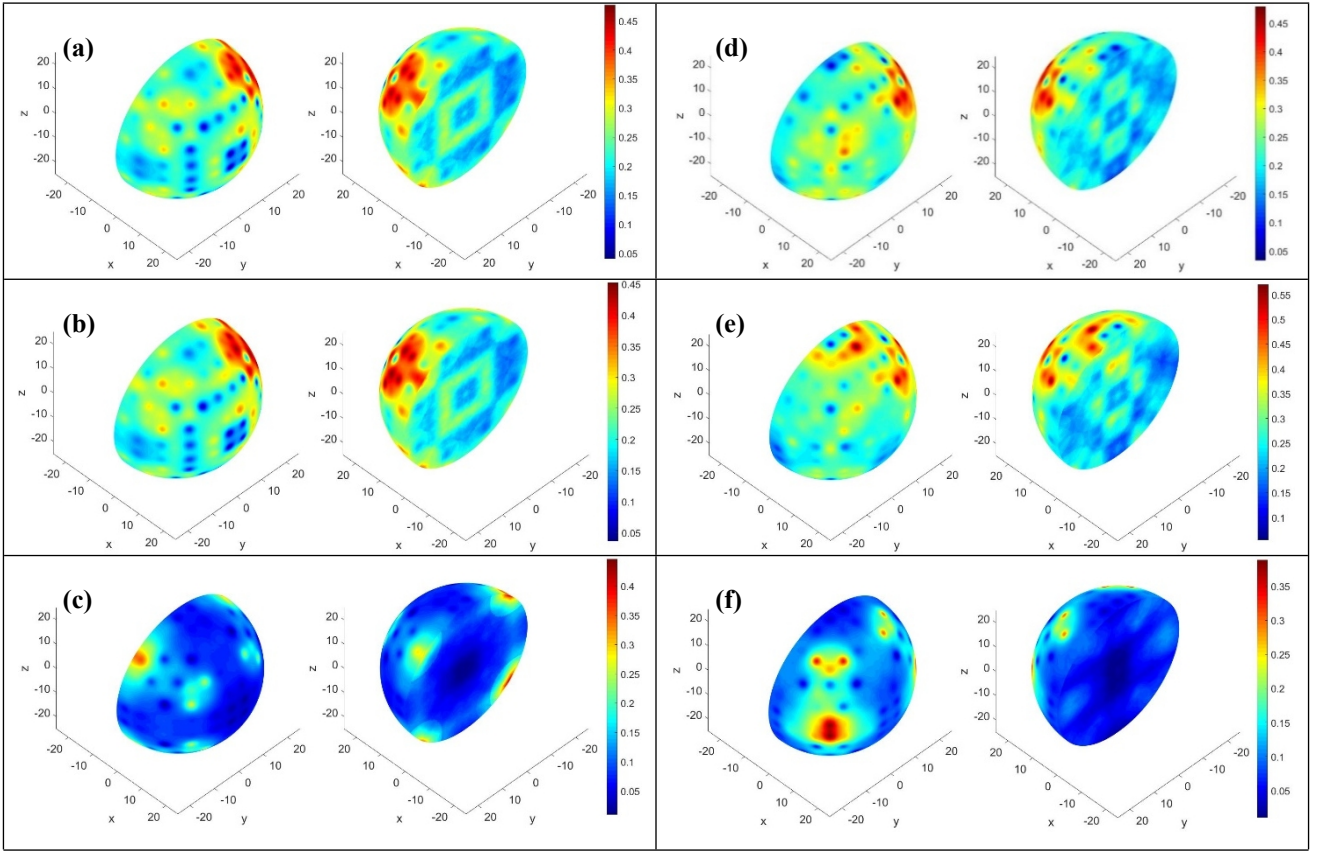

Figure 11. Normalized field intensity plots of Au@Ag illuminated with TE (a-c) and TM (d-f) polarized wave: (a,d) at $\lambda$=370 nm without external tip, (b,e) $\lambda$=370 nm and (c,f) $\lambda$=490 nm with GaP tip.

Considering the results shown in Figures 9 – 12 and Table 3, absorption enhancements are found greater when the GaP tip is used. Other than the enhancement cases, external tip materials can also damp the absorption efficiency values of both bare metallic NPs and bimetallic core-shell pairs as compared to the cases without external tip. Comparing the TE polarized incident wave with and without external tip use shows that the introduction of tip damps the absorption efficiency values of all NP types due to lack of vertical coupling along the tip-NP axis. However, the absorption behavior is different for TM-polarized cases. Bare Ag and Au@Ag core-shell NPs are observed to have multipeak response, where the absorption enhancement occurs at the 2nd LSPR wavelength, $\lambda$=490 nm. It is found that the damping of the absorption at the 1st LSPR wavelength, $\lambda$=370 nm, results from the nonzero imaginary part of the refractive index of the tip material. However, the damping under TM-polarized incident light in presence of an external tip is not observed for bare Au and Ag@Au core-shell NPs since the increasing driving force with the introduction of an external tip can dominate over the damping factor of the NPs. This concludes the effect of shell material on the absorption behavior. Therefore, our study shows consistent results found in literature based on the damping effect and it is further extended to show the dielectric transition of external tip materials.

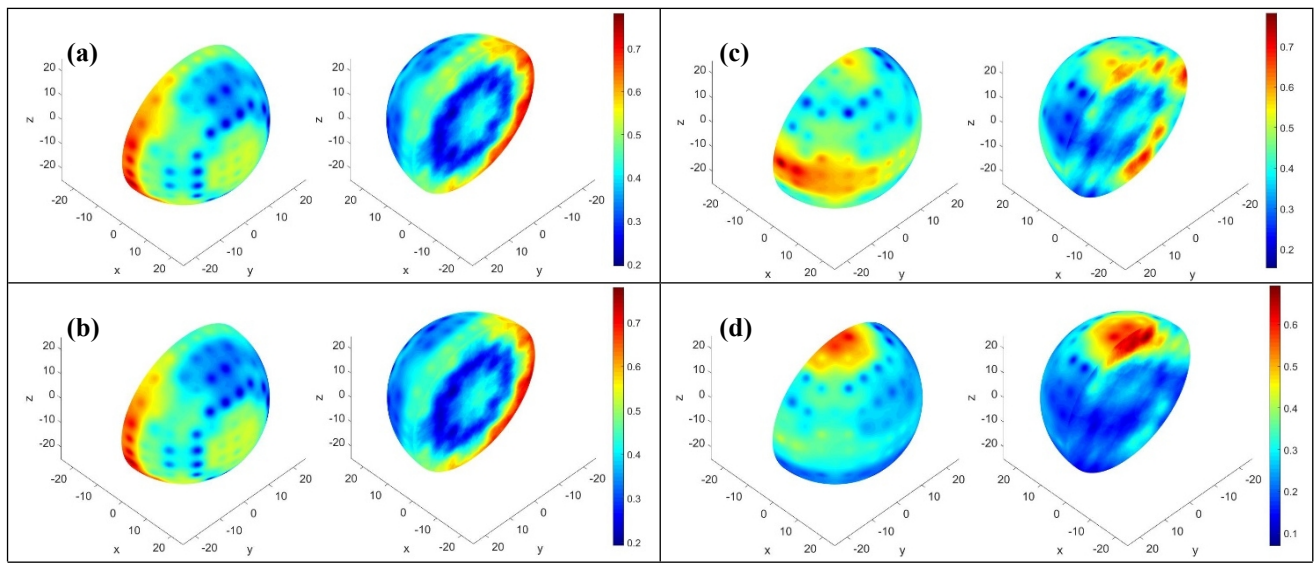

Figure 12. Normalized field intensity plots of Ag@Au with TE (a,b) and TM (c,d) polarized wave at $\lambda$=380 nm: (a,c) without external tip and (b,d) with GaP tip.

## 4. CONCLUSIONS

Spectral absorption profiles and LSPR response of the metallic core-shell nanoparticles are studied to discern their sensitivities to the changes in core/shell size configurations and the dielectric medium properties. The vectorized version of DDA-SI (DDA-SI-v) is used for the numerical analyses of Ag-Au core-shell pairs being placed over a glass substrate. Comparisons of absorption enhancement and LSPR wavelength shifts are performed with size parameter, $f_r$, for fixed sizes of the core and the nanoparticle. More than one plasmon mode can be observed due to the hybrid structure of the core-shell NPs. Plasmon response of the Ag core-Au shell case is observed with blue-shift and red-shift in the LSPR wavelength for the first and secondary peaks, respectively. However, the plasmon response of the Au core-Ag shell case is affected by the shielding of the Ag shell that leads to the greatest absorption only at the extraordinary mode.

Parametric size comparisons showed that increasing volumetric filling ratio, $f_r$, of core gives rise to maximum absorption enhancement over 91% for Ag@Au core-shell NP. However, an increase in $f_r$ of the core in the Au@Ag case results in damping of absorption efficiency values by 135%. These significant differences in absorption can be utilized in the characterization or sensing of the core-shell NPs and core/shell formation applications. Moreover, the changes in the refractive index of the dielectric medium above the glass substrate are compared for the absorption efficiency values and LSPR responses of the fixed core cases for both Ag@Au and Au@Ag core-shell NPs. It is shown that interaction between the metallic shell surface and dielectric medium leads to absorption enhancement as the refractive index of the medium increases. Maximum absorption enhancement is found for the Ag@Au shell case while greater redshift in LSPR wavelength occurs in the Au@Ag shell case. It is shown that there is a trade-off between the absorption enhancement and red-shift in the LSPR response. Therefore, selected core-shell NPs can be effectively used in biosensing and material detection by comparing their LSPR and absorption responses.

Further analyses are performed on the spectral absorption behavior of the selected core-shell (Ag@Au, Au@Ag) and bare metallic (Ag and Au) NPs that are placed under a dielectric external tip. External tip materials are considered with their changing dielectric behavior at increasing wavelengths. The results showed that the external tip introduction can both damp and enhance the absorption of the NPs. The analysis showed that the absorption enhancement with GaP tip was greater. Moreover, extended spectral absorption analyses show that damping characteristics of the external tip use can alter for different shell materials. Using a wide range of spectrum also helped us to observe the dielectric transition of the external tip materials.

Considering the redshifts in LSPR wavelengths, further analysis can be performed by keeping core diameter or shell thickness as constant while increasing the overall nanoparticle size. This study can also be extended further with local heating of an array of core-shell NPs in the presence of an external tip for nanomanufacturing purposes. Moreover, changing the medium refractive index similar to visceral organs can be utilized in tumor treatment applications to simulate the photothermal heating of the NPs under the effect absorption enhancement of an external tip.

**REFERENCES**

[1]     Maier S A 2007 *Plasmonics: Fundamentals and applications* (Springer US)
[2]     Lee Y H, Chen H, Xu Q H and Wang J 2011 Refractive index sensitivities of noble metal nanocrystals: The effects of multipolar plasmon resonances and the metal type *J. Phys. Chem. C* **115** 7997–8004
[3]     Zhu J, Li X, Li J and Zhao J 2018 Enlarge the biologic coating-induced absorbance enhancement of Au-Ag bimetallic nanoshells by tuning the metal composition *Spectrochim. Acta Part A Mol. Biomol. Spectrosc.* **189** 571–7
[4]     Zhu J, Li J and Zhao J 2014 The Study of Surface Plasmon Resonance in Au-Ag-Au Three-Layered Bimetallic Nanoshell : The Effect of Separate Ag Layer *Plasmonics* **9** 435–41
[5]     Sharma R, Roopak S, Alok P and Sharma R P 2017 Study of Surface Plasmon Resonances of Core-Shell Nanosphere : A Comparison between Numerical and Analytical Approach *Plasmonics* **12** 977–86
[6]     Chen Y, Wu H, Li Z, Wang P, Yang L and Fang Y 2012 The Study of Surface Plasmon in Au/Ag Core/Shell Compound Nanoparticles *Plasmonics* **7** 509–13
[7]     Xu X, Liu M, Luo J, Wang Y, Yi Z, Li X, Yi Y and Tang Y 2015 Nanoscale Energy Confinement and Hybridization of Surface Plasmons Based on Skin Depth in Au / Ag Core-Shell Nanostructures *Plasmonics* **10** 797–808
[8]     Zhang C, Chen B, Li Z and Chen Y 2015 Surface Plasmon Resonance in Bimetallic Core − Shell Nanoparticles *J. Phys. Chem. C* **119** 16836–45
[9]     Hubenthal F, Borg N and Träger F 2008 Optical properties and ultrafast electron dynamics in gold-silver alloy and core-shell nanoparticles *Appl. Phys. B Lasers Opt.* **93** 39–45
[10]    Avsar D, Erturk H and Mengüç M P 2019 Plasmonic responses of metallic/dielectric core-shell nanoparticles on a dielectric substrate *Mater. Res. Express* **6**
[11]    Navas M P and Soni R K 2015 Laser-Generated Bimetallic Ag-Au and Ag-Cu Core-Shell Nanoparticles for Refractive Index Sensing *Plasmonics* **10** 681–90
[12]    Hawes E A, Hastings J T, Crofcheck C and Mengüç M P 2008 Spatially selective melting and evaporation of nanosized gold particles. *Opt. Lett.* **33** 1383–5
[13]    Talebi Moghaddam S, Avşar D, Ertürk H and Mengüç M P 2017 Effect of the probe location on the absorption by an array of gold nano-particles on a dielectric surface *J. Quant.*

*Spectrosc. Radiat. Transf.* **197** 106–13

[14] Talebi Moghaddam S, Ertürk H and Mengüç M P 2016 Enhancing local absorption within a gold nano-sphere on a dielectric surface under an AFM probe *J. Quant. Spectrosc. Radiat. Transf.* **178** 124–33

[15] Huda G M, Donev E U, Mengüç M P and Hastings J T 2011 Effects of a silicon probe on gold nanoparticles on glass under evanescent illumination *Opt. Express* **19** 12679

[16] Huda G M and Todd Hastings J 2013 Absorption modulation of plasmon resonant nanoparticles in the presence of an AFM tip *IEEE J. Sel. Top. Quantum Electron.* **19** 4602306

[17] Huda G M, Mengüç M P and Hastings J T 2012 Absorption suppression of silver nanoparticles in the presence of an AFM tip: A harmonic oscillator model *AIP Conf. Proc.* **1475** 134–6

[18] Albella P, Alcaraz de la Osa R, Moreno F and Maier S A 2014 Electric and Magnetic Field Enhancement with Ultralow Heat Radiation Dielectric Nanoantennas: Considerations for Surface-Enhanced Spectroscopies *ACS Photonics* **1** 524–9

[19] Loke V L Y, Mengüç M P and Nieminen T A 2011 Discrete-dipole approximation with surface interaction: Computational toolbox for MATLAB *J. Quant. Spectrosc. Radiat. Transf.* **112** 1711–25

[20] Rostampour Fathi Z, Mengüç M P and Ertürk H 2018 Plasmon coupling between complex gold nanostructures and a dielectric substrate *Appl. Opt.* **57** 8954–63

[21] Friebel M and Meinke M 2006 Model function to calculate the refractive index of native hemoglobin in the wavelength range of 250-1100 nm dependent on concentration *Appl. Opt.* **45** 2838

[22] Smithson Adair G and Elaine Robinson M 1930 The Specific Refraction Increments of Serum-Albumin and Serum-Globulin *Biochem. J.* **24** 993–1011

[23] Sathiyamoorthy K and Kolios M C 2015 Numerical investigation of plasmonic properties of gold nanoshells *Proceedings of SPIE* vol 9340, ed T Vo-Dinh and J R Lakowicz p 93400V

[24] Johnson P B and Christy R W 1972 Optical Constants of the Noble Metals *Phys. Rev. B* **6** 4370–9

[25] Aspnes D E and Studna A A 1983 Dielectric functions and optical parameters of Si, Ge, GaP, GaAs, GaSb, InP, InAs, and InSb from 1.5 to 6.0 eV *Phys. Rev. B* **27** 985–1009

[26] Palik E D 1998 *Handbook of optical constants of solids* (Academic Press)